\documentclass[]{aastex631}
\usepackage{amsmath}

\shorttitle{}
\graphicspath{{./}{figures/}}

\begin{document}

\title{Anticipating the DART impact: Orbit estimation of Dimorphos using a simplified model}

\author{Shantanu P. Naidu}
\affiliation{Jet Propulsion Laboratory, California Institute of Technology}

\author{Steven R. Chesley}
\affiliation{Jet Propulsion Laboratory, California Institute of Technology}

\author{Davide Farnocchia}
\affiliation{Jet Propulsion Laboratory, California Institute of Technology}

\author{Nick Moskovitz}
\affiliation{Lowell Observatory}

\author{Petr Pravec}
\affiliation{Astronomical Institute of the Czech Academy of Sciences}

\author{Petr Scheirich}
\affiliation{Astronomical Institute of the Czech Academy of Sciences}

\author{Cristina Thomas}
\affiliation{Northern Arizona University}

\author{Andrew S. Rivkin}
\affiliation{Johns Hopkins University, Applied Physics Laboratory}

\begin{abstract}
We used the times of occultations and eclipses between 
the components of the 65803 Didymos binary system observed in its
lightcurves from 2003-2021 to estimate the
orbital parameters of Dimorphos relative to Didymos.
We employed a weighted least-squares approach and a modified 
Keplerian orbit model in order
to accommodate the effects from non-gravitational forces 
such as Binary YORP that could cause
a linear change in mean motion over time. We estimate that the period of the mutual 
orbit at the epoch 2022 September 26.0 TDB, the day of the DART impact, is 
$11.9214869 \pm 0.000028$~h ($1\sigma$) and that the
mean motion of the orbit is changing at a rate of 
$(5.0\pm 1.0)\times 10^{-18}$~rad s$^{-2}$ $(1\sigma$). 
The formal $3\sigma$ uncertainty in orbital phase of Dimorphos 
during the planned Double Asteroid Redirection Test (DART) 
mission is $5.4^\circ$. Observations from July to September 2022, 
a few months to days prior 
to the DART impact, should provide modest improvements to the orbital 
phase uncertainty and reduce it
to about $4.2^\circ$. 
These results, generated using a relatively simple model, are consistent with those 
generated using the more sophisticated model of
\citet{scheirich22}, which demonstrates the reliability of our method and adds
confidence to these mission-critical results.

\end{abstract}


\section{Introduction} \label{sec:intro}

The binary near-Earth asteroid (65803) Didymos is the target of
NASA's Double Asteroid Redirection Test (DART)
mission, a test of the kinetic impactor approach to planetary defense~\citep{cheng16}.
The mission was launched on 2021 November 22 and the spacecraft will impact 
the satellite of Didymos, named Dimorphos, on 2022 September 26. 
The primary 
objective of the DART mission is to change the orbital period of 
Dimorphos and measure this change using ground-based observations~\citep{rivkin21}. 
The DART spacecraft carried along with it a Light Italian Cubesat for Imaging of Asteroids (LICIA cube), built by the Italian Space Agency, for observations of
the Didymos system during the DART impact~\citep{dotto21}. Didymos is also the target of
the European Space Agency's proposed Hera mission, which will rendezvous several years
after DART for post-impact characterization of the target~\citep{michel18}.

Didymos was discovered in 1996 by the Spacewatch telescope at Kitt Peak (MPEC 
1996-H02)\footnote{https://www.minorplanetcenter.net/mpec/J96/J96H02.html} and its binary
nature was discovered in 2003 November by
\citet{pravec03} when mutual events (occultations/eclipses) 
were observed in lightcurves. 
The presence of a satellite was confirmed later that month when Arecibo radar images
resolved echoes from the two objects~\citep{naidu20}. \citeauthor{naidu20} used the radar and lightcurve data from 2003 to 
characterize the physical properties of the system, including a 3D shape model of the
primary, size of the secondary, and mutual orbit parameters. 
The primary and secondary components are roughly
780 m and 150 m in diameter, respectively, and the mutual orbit of the
system has a semimajor axis of $\sim1.2$~km and a period of about
11.9~h~\citep{pravec06,naidu20}.

In order to characterize the pre-impact orbit of Dimorphos, ~\citet{pravec22} obtained
lightcurves of the system in 2015, 2017, 2019, 2020, and 
2021 in addition to the original lightcurves from 2003~\citep{pravec06}.
\citet{scheirich22} use the primary-subtracted lightcurves (where the
primary lightcurve has been modeled and subtracted from the total lightcurve)
to fit a binary asteroid model, including the mutual orbit parameters. They 
model the binary system as two ellipsoids orbiting each other on a 
modified Keplerian orbit. They use ray-tracing code and photometric 
models such as Lommel-Seeliger and Lambert scattering laws
to model the orbital lightcurves. 
Similar methods were used to model orbits of other binary asteroids~\citep[e.g.][]{scheirich09}.

In this paper we develop a simpler approach to 
estimate the mutual orbit parameters
of the system by using only the times of the beginnings and ends of 
mutual events. This approach differs from that of \citet{scheirich22} 
in that it uses a different observational data type and different
observational model. Our simplified approach allows us to determine the orbital parameters quickly compared to the approach of \citet{scheirich22} and our method is robust as demonstrated by the consistency of the results with those of \citet{scheirich22}.

\section{Observations}

We used lightcurves
from \citet{pravec06} and \citet{pravec22} and 
measured the times of mutual events 
detected in the orbital component 
of the lightcurves of Didymos. 
The decomposition of the total lightcurve into the primary and the
primary-subtracted components was done by \citet{pravec06} for 
the 2003 data, by \citet{pravec22}
for the 2015, 2017, and 2019 data, 
and by one of us (NM) for the 2020 and 2021 data. 
We used the \citet{scheirich22} model to determine
the types of events. We could have determined the types 
of events without using the \citet{scheirich22} model, 
but it would have involved trial and error
and involved additional effort. There are four
different kinds of mutual events in the data: eclipse of the primary
(secondary casting shadows on the primary), eclipse of the secondary,
occultation of the primary from the point of view of the observer, and
occultation of the secondary. 

A mutual event causes a drop in the brightness of the system and
typically has four contact times: the first contact is when the event
begins and the brightness starts decreasing, the second contact is
when the brightness reaches a minimum, the third contact is when the
brightness starts increasing again, and the fourth contact is when the
event ends and the brightness of the system 
returns to the baseline value. We use $T_1$, $T_2$, $T_3$, and $T_4$ to refer to these
contact times.
If the Sun-asteroid-observer phase
angle is low, an eclipse and an
occultation might overlap such that the event 
could begin as an eclipse/occultation and 
end as an occultation/eclipse. In some cases we might have
partial events in which only a fraction
of the primary and secondary undergo a mutual event.
Overlapping events have ill-defined 
$T_2$ and $T_3$ whereas partial events lack $T_2$ and $T_3$. 
Complete secondary events have
consistent depths because the entire contribution from the secondary 
vanishes, whereas the depths of complete primary events vary 
because the brightness of the primary varies across its surface. 
We neglect possible albedo variations on the primary, 
as well as phase effects at higher phase angles, 
both of which can affect the apparent local surface brightness 
of the primary. These assumptions imply that the primary 
event depths match the secondary event depths.
We measured event times $T_{1.5}$ and $T_{3.5}$ 
(for both primary and secondary events) 
as the times when
the drop in brightness of the system was half of the total drop
in brightness due to a full secondary event. 
For a typical non-overlapping and 
complete event $T_{1.5}$ lies between $T_1$ and $T_2$, 
and $T_{3.5}$ lies between $T_3$ and $T_4$. 
We used $T_{1.5}$ and $T_{3.5}$ 
as our observations for the orbit determination.

\begin{figure}
        \centering      
        \includegraphics[width=\textwidth]{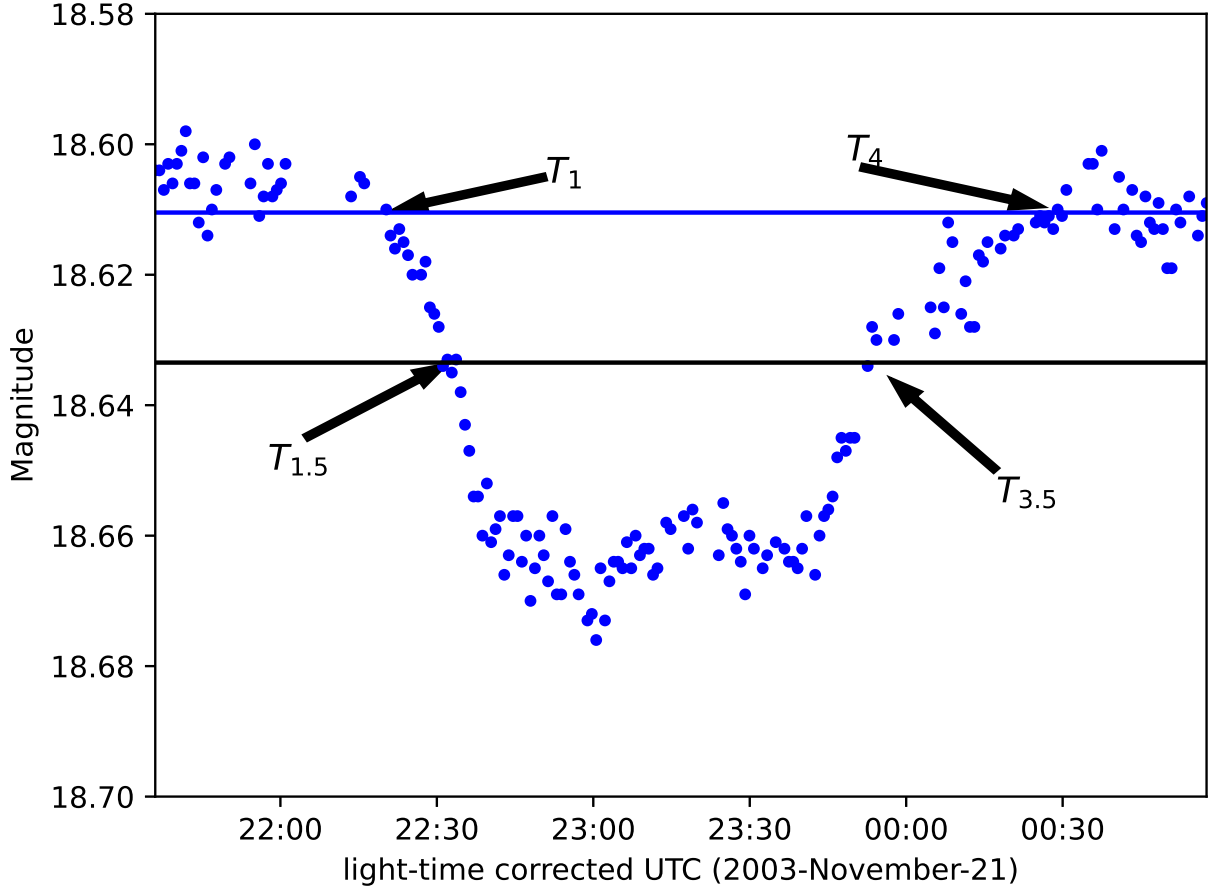}
        \caption{A secondary eclipse event showing various contact times. 
        The horizontal blue line represents the baseline magnitude of the system 
        and is computed by taking the average magnitude of the points just 
        outside the mutual event. The black line indicates a magnitude that 
        is fainter than the baseline by 0.023 mag, which represents a 2.1\% drop 
        in brightness of the system.}
        \label{fig:event}
\end{figure}

For the Didymos system, with a diameter ratio (secondary/primary) of 0.21,
a full secondary event causes a drop in brightness of
4.2\%~\citep{pravec06}\footnote{The latest estimate of the secondary event depth is 4.5\%~\citep{scheirich22}. 
The work in this paper was done
before the latest estimate was available, however our assigned measurement
uncertainties accommodate this difference.}. We measured
$T_{1.5}$ and $T_{3.5}$ as the times when
the brightness drops by 2.1\%, 
which corresponds to a magnitude increase of 0.023. 
When the Sun-target-observer phase angle is zero, $T_{1.5}$ and $T_{3.5}$
correspond to start and end times of when the center-of-mass of the satellite undergoes a mutual event.
Figure~\ref{fig:event} shows the contact times for a secondary eclipse event from 2003.
We used the average magnitude of the points adjacent to the event as the baseline magnitude of the lightcurve outside each event
and plotted a horizontal line 0.023 magnitude above this baseline. We made measurements
visually by inspecting the intersection of this line with the mutual events.
We assigned $1\sigma$ uncertainties of $(T_{1.5}-T_1)/2$ and $(T_4 - T_{3.5})/2$
to $T_{1.5}$ and $T_{3.5}$, respectively.
The assigned uncertainties take into account measurement errors as well as errors and variations in the assumed event depths. Table~\ref{tab:observations} lists all the observations. 

\newpage
\begin{longtable}{llllllr}
  \caption{Mutual event times measured in observations from 2003 to 2021. 
    All times are one-way light-time corrected to reflect 
    the time of the events 
    at the asteroid, not the times that they were observed from Earth.
    The last column shows the residuals (observed - computed) 
    for solution 104, which is described in Section~\ref{sec:result}}\\*
                                       &                     &              &  Occulted/Eclipsed  &                    & $1\sigma$ Uncertainty & Residuals\\*
    Calendar date (UTC)  &  Julian date & Contact &  object                     & Event type  & (days)  & (sigmas)            \\*
    \hline  
2003 Nov 20 22:48:00 & 2452964.4500 & 1.5 & Secondary & Eclipse & 0.004  &             0.7177 \\
2003 Nov 21 00:01:26 & 2452964.5010 & 3.5 & Secondary & Eclipse & 0.004  &  	      -0.5260 \\
2003 Nov 21 22:31:00 & 2452965.4382 & 1.5 & Secondary & Eclipse & 0.005  &  	      -0.1701 \\
2003 Nov 21 23:52:30 & 2452965.4948 & 3.5 & Secondary & Eclipse & 0.011  &            -0.0028 \\  
2003 Nov 22 04:32:35 & 2452965.6893 & 1.5 & Primary & Eclipse & 0.004  &  	       0.5784 \\
2003 Nov 22 05:50:21 & 2452965.7433 & 3.5 & Primary & Eclipse & 0.004  &  	       0.1697 \\
2003 Nov 23 04:19:46 & 2452966.6804 & 1.5 & Primary & Eclipse & 0.004  &  	       0.3414 \\
2003 Nov 23 05:38:49 & 2452966.7353 & 3.5 & Primary & Eclipse & 0.003  &  	       0.2542 \\
2003 Nov 24 04:03:12 & 2452967.6689 & 1.5 & Primary & Eclipse & 0.003  &  	      -0.7358 \\
2003 Nov 24 05:26:52 & 2452967.7270 & 3.5 & Primary & Eclipse & 0.006  &  	       0.0447 \\
2003 Nov 26 03:40:36 & 2452969.6532 & 1.5 & Primary & Eclipse & 0.004  &  	      -0.5276 \\
2003 Nov 26 05:03:33 & 2452969.7108 & 3.5 & Primary & Eclipse & 0.003  &  	      -0.0019 \\
2003 Nov 27 21:27:12 & 2452971.3939 & 1.5 & Secondary & Eclipse & 0.005  &  	       0.4601 \\
2003 Nov 29 21:01:17 & 2452973.3759 & 1.5 & Secondary & Eclipse & 0.003  &  	       0.0022 \\
2003 Nov 30 02:57:33 & 2452973.6233 & 1.5 & Primary & Eclipse & 0.003  &  	      -0.1950 \\
2003 Dec 02 03:55:35 & 2452975.6636 & 3.5 & Primary & Occultation & 0.005  &  	      -0.3090 \\
2003 Dec 03 03:38:44 & 2452976.6519 & 3.5 & Primary & Occultation & 0.011  &  	      -0.4991 \\
2003 Dec 03 08:16:13 & 2452976.8446 & 1.5 & Secondary & Eclipse & 0.006  &  	      -0.6084 \\
2003 Dec 03 09:46:22 & 2452976.9072 & 3.5 & Secondary & Occultation & 0.009  &         0.1859 \\
2003 Dec 04 02:17:31 & 2452977.5955 & 1.5 & Primary & Eclipse & 0.004  &  	       0.6318 \\
2003 Dec 04 03:35:59 & 2452977.6500 & 3.5 & Primary & Occultation & 0.004  &  	       0.0754 \\
2003 Dec 18 23:29:19 & 2452992.4787 & 1.5 & Primary & Eclipse & 0.013  &  	       0.0933 \\
2003 Dec 19 00:50:58 & 2452992.5354 & 3.5 & Primary & Occultation & 0.009  &  	      -0.2509 \\
2003 Dec 19 05:23:16 & 2452992.7245 & 1.5 & Secondary & Eclipse & 0.009  &  	      -0.1319 \\
2003 Dec 19 06:44:55 & 2452992.7812 & 3.5 & Secondary & Occultation & 0.008  &        -0.6452 \\
2003 Dec 20 05:19:06 & 2452993.7216 & 1.5 & Secondary & Eclipse & 0.004  &  	       0.8909 \\
2003 Dec 20 06:32:15 & 2452993.7724 & 3.5 & Secondary & Occultation & 0.008  &        -0.8477 \\
\hline
2015 Apr 13 04:54:20 & 2457125.7044 & 3.5 & Primary & Occultation & 0.007  &  	       0.0336 \\
2015 Apr 14 09:25:37 & 2457126.8928 & 1.5 & Secondary & Eclipse & 0.004  &  	      -0.6243 \\
\hline
2017 Feb 25 03:50:06 & 2457809.6598 & 1.5 & Primary & Occultation & 0.005  &  	      -0.3512 \\
2017 Feb 25 05:45:10 & 2457809.7397 & 3.5 & Primary & Eclipse & 0.006  &  	       1.6079 \\
2017 Apr 18 07:46:16 & 2457861.8238 & 1.5 & Primary & Eclipse & 0.004  &  	       0.0144 \\
2017 May 04 06:49:32 & 2457877.7844 & 3.5 & Primary & Occultation & 0.005  &  	       0.3233 \\
\hline
2019 Jan 31 08:39:24 & 2458514.8607 & 3.5 & Secondary & Eclipse & 0.007  &  	      -0.5171 \\
2019 Jan 31 13:03:21 & 2458515.0440 & 1.5 & Primary & Occultation & 0.004  &  	       0.6895 \\
2019 Mar 09 01:42:31 & 2458551.5712 & 1.5 & Secondary & Occultation & 0.007  &         0.4536 \\
2019 Mar 09 02:35:13 & 2458551.6078 & 3.5 & Secondary & Eclipse & 0.005  &  	      -0.0914 \\
2019 Mar 10 02:15:47 & 2458552.5943 & 3.5 & Secondary & Eclipse & 0.005  &  	      -1.2862 \\
2019 Mar 11 02:15:30 & 2458553.5941 & 3.5 & Secondary & Eclipse & 0.005  &  	      -0.0786 \\
\hline
2020 Dec 12 14:21:50 & 2459196.0985 & 1.5 & Primary & Occultation & 0.004  &  	       0.9640 \\
2020 Dec 12 15:00:34 & 2459196.1254 & 3.5 & Primary & Occultation & 0.004  &  	       0.1897 \\
2020 Dec 17 08:50:38 & 2459200.8685 & 1.5 & Secondary & Eclipse & 0.006  &  	       0.8257 \\
2020 Dec 17 09:36:43 & 2459200.9005 & 3.5 & Secondary & Eclipse & 0.006  &  	      -0.3763 \\
2020 Dec 23 08:32:55 & 2459206.8562 & 3.5 & Secondary & Eclipse & 0.007  &  	      -0.3514 \\
2020 Dec 23 12:35:42 & 2459207.0248 & 1.5 & Primary & Occultation & 0.006  &  	       1.0714 \\
2020 Dec 23 13:04:47 & 2459207.0450 & 3.5 & Primary & Occultation & 0.007  &  	      -0.3992 \\
\hline
2021 Jan 08 10:57:12 & 2459222.9564 & 1.5 & Primary & Eclipse & 0.005  &  	       0.4413 \\
2021 Jan 08 11:35:39 & 2459222.9831 & 3.5 & Primary & Eclipse & 0.006  &  	      -0.7766 \\
2021 Jan 09 10:50:26 & 2459223.9517 & 1.5 & Primary & Eclipse & 0.010  &  	       0.4727 \\
2021 Jan 09 11:21:15 & 2459223.9731 & 3.5 & Primary & Eclipse & 0.010  &  	      -0.7482 \\
2021 Jan 10 11:11:02 & 2459224.9660 & 3.5 & Primary & Eclipse & 0.007  &  	      -1.0587 \\
2021 Jan 12 15:06:46 & 2459227.1297 & 1.5 & Secondary & Occultation & 0.013  &        -0.1800 \\
2021 Jan 14 09:56:44 & 2459228.9144 & 1.5 & Primary & Eclipse & 0.004  &  	       0.7749 \\
2021 Jan 14 10:26:41 & 2459228.9352 & 3.5 & Primary & Eclipse & 0.009  &  	      -1.0657 \\
2021 Jan 17 14:26:00 & 2459232.1014 & 1.5 & Secondary & Occultation & 0.007  &        -0.0537 \\
2021 Jan 18 08:22:50 & 2459232.8492 & 1.5 & Primary & Occultation & 0.011  &  	      -0.0715 \\
2021 Jan 18 09:00:17 & 2459232.8752 & 3.5 & Primary & Occultation & 0.007  &  	       0.1228 \\
2021 Jan 18 09:14:41 & 2459232.8852 & 1.5 & Primary & Eclipse & 0.007  &  	       0.4403 \\
2021 Jan 18 09:40:45 & 2459232.9033 & 3.5 & Primary & Eclipse & 0.017  &  	      -0.7859 \\
2021 Jan 20 01:54:20 & 2459234.5794 & 1.5 & Secondary & Occultation & 0.006  &        -1.2584 \\
2021 Jan 20 03:40:19 & 2459234.6530 & 3.5 & Secondary & Eclipse & 0.011  &  	      -0.1952 \\
\hline
\label{tab:observations}
\end{longtable}

In addition to lightcurve mutual events, 
radar range and Doppler measurements of Dimorphos relative to Didymos 
were also available from Table 6 of \citet{naidu20}. 
We did not include these in the final orbit fit because they only span a short period 
in 2003 and do not provide any significant constraints to the orbital uncertainties
during the DART impact in 2022 September. However, we used the
radar measurements to check for consistency with the best-fit orbit 
by generating range and Doppler predictions at the time of the observations.

\section{Orbit fit}

We used a weighted least-squares method to estimate the best-fit model
parameters~\citep{milani09}. The goal is to minimize the cost function, $\chi^2 =
\boldsymbol\nu^T W \boldsymbol\nu$, where $\boldsymbol\nu$ is the array of residuals (observed -
computed) and $W$ is the weight matrix with $W_{i,j} = 0$ for $i \neq
j$, and $W_{i,j}=1/\sigma_i^2$ for $i=j$, and $\sigma_i$ is the
observational uncertainty for the $i$th observation.

The least-squares solution is found by iteratively
correcting the estimated parameters $\boldsymbol x$ by
\begin{equation}
 \Delta \boldsymbol x = - \Gamma B^T W\boldsymbol\nu,
\end{equation}
where $B = \partial \boldsymbol\nu/\partial x$ is the design matrix,
$\Gamma = C^{-1}$ is the covariance matrix, and $C=B^T W B$ is the
normal matrix, also called the information matrix. This iterative procedure is
called differential corrections. The marginal $1\sigma$ uncertainties
of the parameters are computed by taking the square root of the
diagonal elements of the covariance matrix.

In order to compute $\boldsymbol\nu$, we have to use a model
to calculate the ``computed" value 
corresponding to each observation. 
We used the NAIF SPICE geometry finder tools~\citep{acton18} 
for this purpose. This calculation requires
SPICE kernels that describe the trajectory, size, shape, and
orientation of the objects. We modeled the primary as an oblate
spheroid with dimensions of 830 x 830 x 786 m~\citep{naidu20} with its
spin pole aligned with the mutual orbit pole. This information is
defined in a planetary constants kernel (PCK) file. The primary was
treated as an ellipsoid for computing the mutual event timings but was
treated as a point mass for computing its gravitational force on the satellite.

We assumed the satellite to be a point mass on a modified Keplerian
mutual orbit around the primary. In addition to Keplerian motion, we
included an additional term for modeling the drift in mean motion due
to Binary YORP~\citep[BYORP,][]{cuk07}. Assuming that the system mass is
constant, a drift in mean motion leads to a change in semimajor axis
with time. The mean anomaly ($M$) and mean motion ($n$) 
of the satellite at time $t$ are given by:
\begin{equation}
  \begin{aligned}
  M(t) = M_0 + n_0(t-t_0) + \frac{1}{2}\dot{n}(t-t_0)^2,\\
  n(t) = n_0 + \dot{n}(t-t_0),\\
  \end{aligned}
  \label{eq:model}
\end{equation}
where $M_0$ and $n_0$ are the mean anomaly and mean motion
of the satellite at time $t_0$, and $\dot{n}$ is
the constant rate of change of mean motion due to BYORP~\citep{cuk07}.
We used these
equations to generate the states of the satellite with
respect to the primary at 1-day intervals and 
saved them as SPK\footnote{SPK
is a file format used by SPICE 
for storing and retrieving ephemeris data. 
These files are required by various 
SPICE routines such as the geometry finder tools that we used for 
calculating the times of occultations
and eclipses.} files.
We used type-5 SPKs, which assume Keplerian motion for
interpolating states. The time interval between states is small
enough that errors in mean motion due to BYORP are orders of magnitude
smaller than the uncertainty. Tests with 0.001 day intervals yield
almost identical results.

To calculate the computed event times, 
we first computed time intervals for mutual events assuming a point-sized satellite. 
When the Sun-target-observer phase angle is zero, these times would
correspond to the observed times, $T_{1.5}$ and $T_{3.5}$. 
However, at non-zero phase angles, eclipses and occultations are 
observable in the lightcurves for
shorter durations: eclipses are observable only when a shadow 
is cast on the part of the target's surface that is visible from Earth, 
while occultations are observable only when the sunlit part of 
the target's surface is occulted. We took these phase effects into account in the 
following way.

For primary events, we computed the point on the surface 
of the primary that is being eclipsed or occulted by the secondary, 
which is assumed to be a point. We then use the SPICE geometry 
finder~\citep{acton18} to calculate 
intervals when the eclipsed point is visible from the Earth 
or when this occulted point is sunlit. The beginnings and ends of these 
intervals are taken to be the computed values of $T_{1.5}$ and $T_{3.5}$. 
Appropriate light-time 
corrections were made using the SPICE routines~\citep{acton18} 
so that the calculated event times correspond to the 
times at the asteroid, which is the 
timescale used in the observed lightcurves.

There are similar phase effects for secondary events. The dark part 
of the surface of the secondary does not contribute to the lightcurves.
The only portion of the secondary surface that contributes to the 
lightcurves is the area that is sunlit and oriented towards Earth.
So, for secondary events, the measured $T_{1.5}$ represents 
the instant when half of this visible area goes 
into an eclipse/occultation and $T_{3.5}$ 
represents the instant when half of the visible area comes out of an eclipse/occultation. 
We corrected the zero-phase mutual event times by computing the
separation between the center of figure and the center 
of the visible area of the secondary and multiplying this separation by the relative 
velocity of the secondary in the direction of the separation. 

For calculating the ``computed" values of radar range separations, 
we used SPICE to subtract the distance of Didymos relative to Earth from 
the distance of Dimorphos relative to Earth.
For calculating Doppler separations ($\Delta f$) 
we computed the magnitude of the 
velocity of Dimorphos relative to Didymos in the direction of Earth ($\Delta v$).
Then Doppler separation was computed as: 
\begin{equation} 
  \Delta f = 2\frac{\Delta v}{c}F
\end{equation}
where $c$ is the speed of light, $F$ is the frequency of radar waves
(8560 MHz for Goldstone and 2380 MHz for Arecibo). 
The factor of two exists due to the fact
that the radar signal gets Doppler shifted twice, once during transmission and once during reception~\citep{ostro93}.

The design matrix, $\partial \boldsymbol\nu/\partial x$, was computed
numerically using second order central differences:
\begin{equation}
  \frac{\partial \boldsymbol\nu}{\partial P} =
  \frac{- \boldsymbol\nu(P+2\delta P) + 8\boldsymbol\nu(P+\delta P)
    - 8\boldsymbol\nu(P- \delta P) + \boldsymbol\nu(P-2 \delta P)}{12 \delta P},
  \label{eq:partials}
\end{equation}
where $\delta P$ is a small increment in the value of the parameter, $P$.
The values for $\delta P$ were carefully chosen by numerically testing
the values of the partials. For $M_0$, $n_0$, and $\dot{n}$ the increments were 0.01 rad, 
$10^{-10}$ rad~s$^{-1}$, and $5\times10^{-18}$~rad~s$^{-2}$.

Our estimable parameters were $M_0$, $n_0$, and $\dot{n}$, whereas
the remaining orbital parameters were fixed. The semimajor axis and eccentricity were set to 1.2 km and 0 respectively, 
based on estimates from \citet{pravec06} and \citet{naidu20}.
The orbit pole longitude and latitude with respect to the IAU76 ecliptic
frame~\citep{seidelmann76} were
set to $320.6^\circ$ and $-78.6^\circ$ respectively, 
based on \citet{scheirich22}. 
These correspond to a longitude of ascending node of 
$50.6^\circ$ and inclination of $168.6^\circ$. 
Given the zero eccentricity, we set the argument of pericenter to zero   
so that the mean anomaly is measured from the ascending node. 

We used an initial value for $n_0$ corresponding to a 
$11.9216$~h orbital period  based on the 
estimate from Scheirich and Pravec (personal communication). We set the 
initial value of $\dot{n}$ to 0 and for $M_0$, we tried initial values 
from $0^\circ$ to $350^\circ$ with a step size of $10^\circ$.

\section{Predictions}

The nominal values of $M$, $n$, and $\dot{n}$ are propagated to a time
$t$ using equations~\ref{eq:model}. The covariance matrix
is mapped to a different epoch using~\citet{milani09}:
\begin{equation}
  \Gamma_t = S \Gamma_0 S^T
\end{equation}
where
\begin{equation*}
  S = \frac{\partial (M_t, n_t, \dot{n})}{\partial (M_0, n_0, \dot{n})} = 
  \begin{bmatrix}
    1       & (t-t_0)      & \frac{1}{2}(t-t_0)^2 \\
    0      &  1             & (t-t_0)                     \\
    0      &  0             & 1                             \\
  \end{bmatrix}
\end{equation*}

Here subscripts $t$ and $0$ denote parameters at time $t$ and $t_0$
respectively. The marginal $1\sigma$ uncertainties on the parameters
are the square roots of the corresponding diagonal elements of
$\Gamma_t$. The uncertainty on $\dot{n}$ does not change with time.

Similarly, predictions of observable uncertainties at time $t$ are
computed as:
\begin{equation}
\Gamma_{T_*} = Q \Gamma_t Q^T,
\end{equation}
where
\begin{equation}
Q = \frac{\partial T_*}{\partial (M_t, n_t, \dot{n})}. 
\end{equation}

Here $T_*$ is the observed event time and $Q$ is computed numerically using second order central
differences.

\section{Results} \label{sec:result}
\begin{figure}
  \centering
  \includegraphics[width=0.7\textwidth]{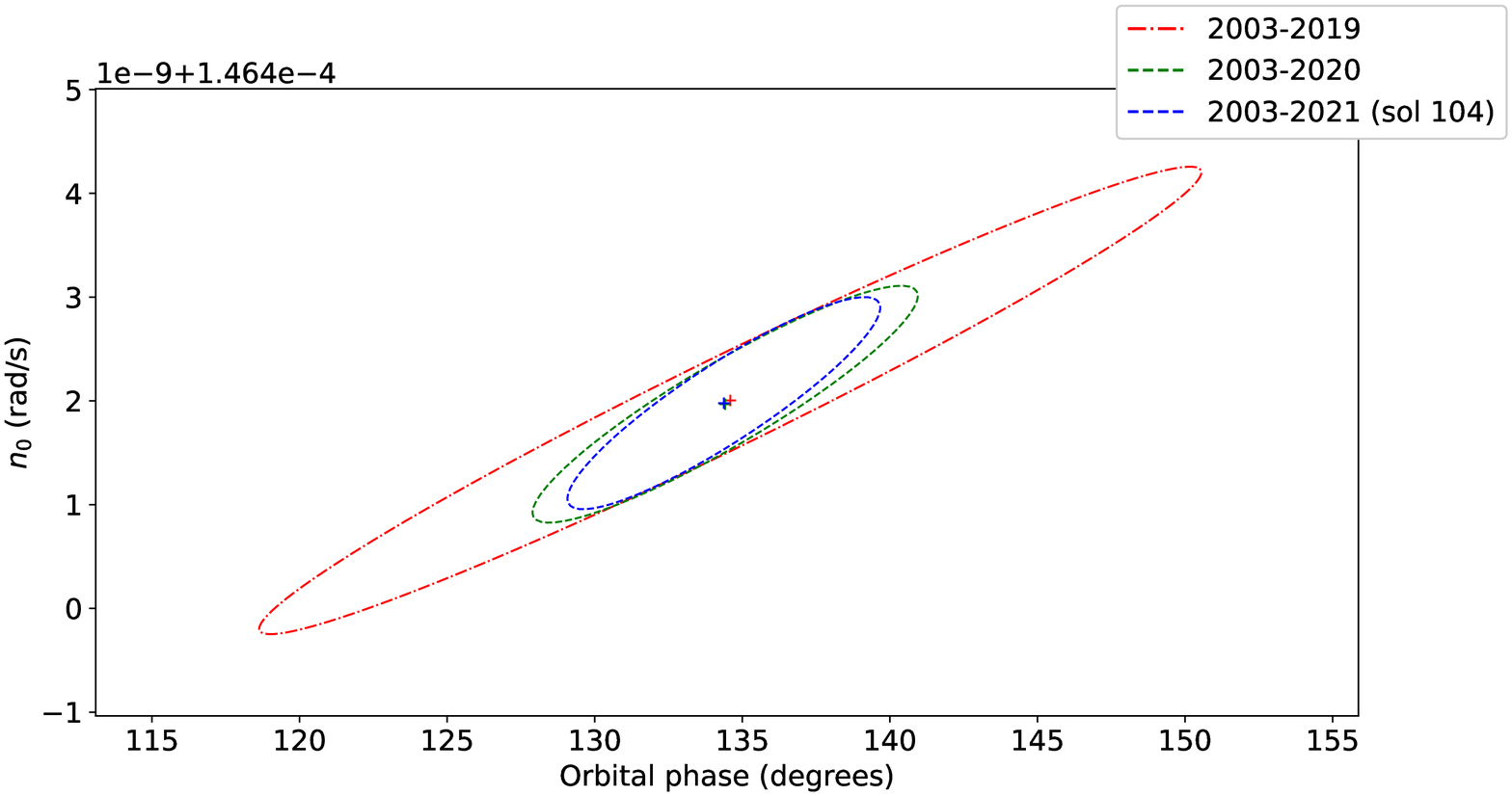}
  \includegraphics[width=0.7\textwidth]{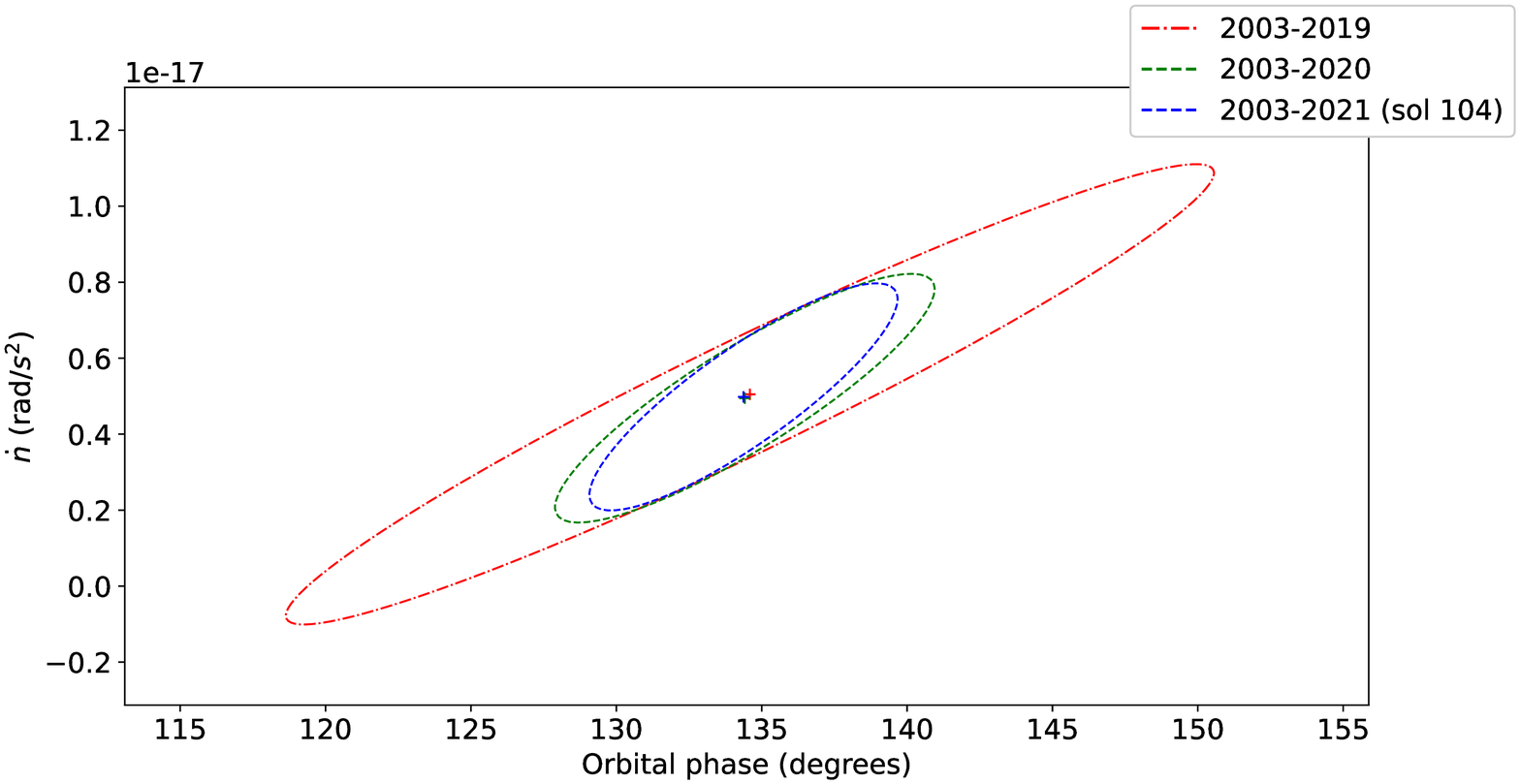}
  \includegraphics[width=0.7\textwidth]{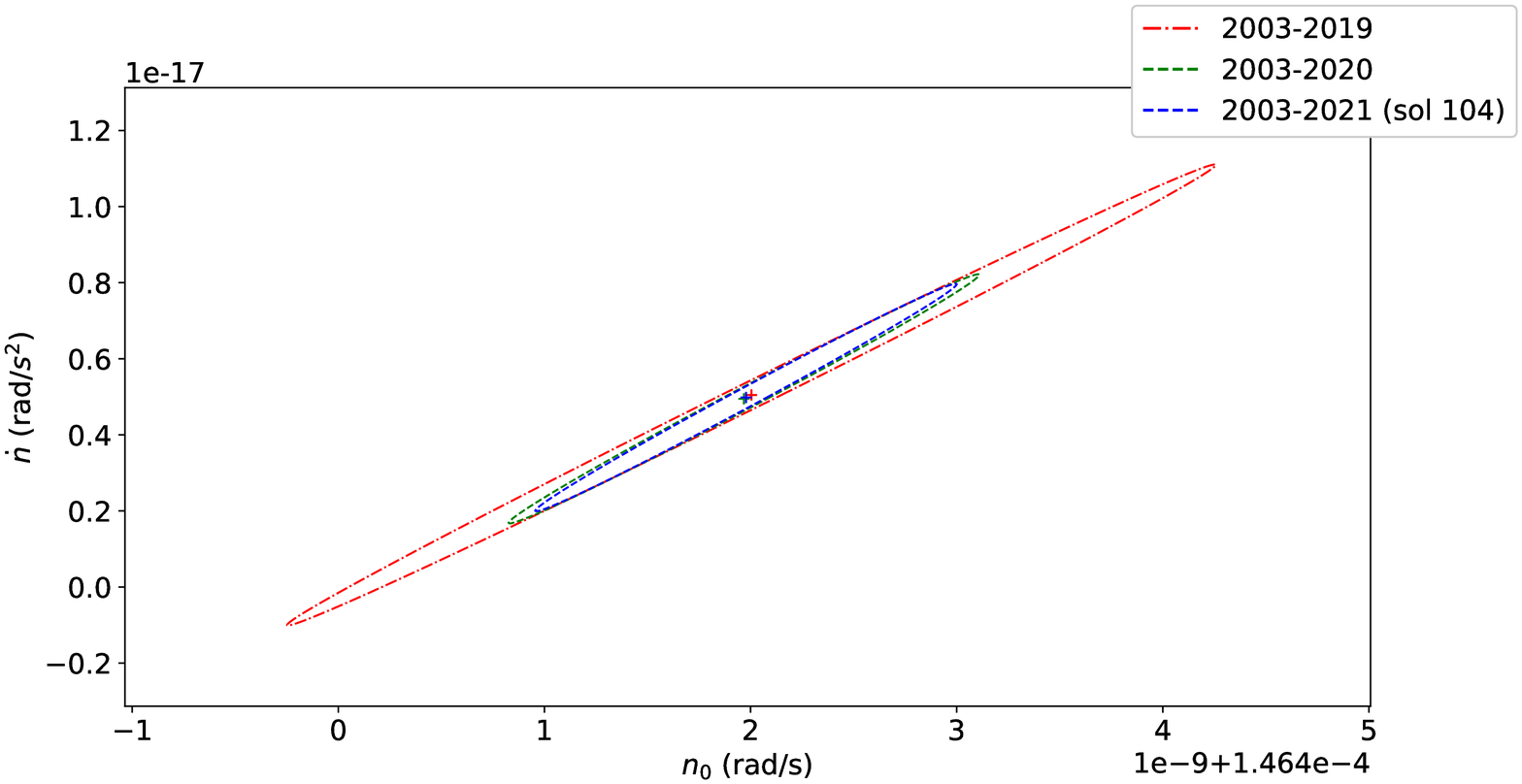}
  \caption{Projections of the best-fit parameters
    and their $3\sigma$ uncertainties 
    for various orbital solutions of Dimorphos.
    The solution using a data-arc 
    of 2003-2019 is shown in red and the solution using
    a data-arc of 2003-2020 is shown in green. 
    Solution 104, the current best-fit solution, 
    is shown in blue. Solutions 
    are mapped to epoch 2022-Sep-26 23:15 UTC, 
    the DART impact time. Orbital phase is the angle in the orbital
    plane measured from the $0^\circ$ 
    longitude in the IAU76 ecliptic 
    frame, as opposed to mean anomaly,
    which is measured from the ascending node. 
    Both angles are measured in the direction of the
    orbital motion of the satellite.}
  \label{fig:solcomp}
\end{figure}

The orbit fits starting from various initial values of $M_0$ converged to 
a single clear best-fit solution. Table~\ref{tab:bestfit} shows the 
best-fit parameters and formal $1\sigma$ uncertainties,
and Table~\ref{tab:cov} shows the corresponding
covariance. Residuals are given in Table~\ref{tab:observations}. The 
value of $\chi^2_\nu$ and the residuals suggest that the 
assigned measurement uncertainties are conservative. 
The estimates from Table~\ref{tab:bestfit} 
are consistent with those from 
\citet{scheirich22} to well within $1\sigma$. 
We refer to this solution as 104. We also 
generated a separate solution
by estimating all the mutual orbit parameters, which yielded 
an orbit pole of 
$(\lambda, \beta) = (310 \pm 15, -76 \pm 4)^\circ$ ($3\sigma$ uncertainties).
This is consistent with the estimate of 
$(\lambda, \beta) = (320.6 \pm 13.7, -78.6 \pm 1.8)^\circ$ 
from \citet{scheirich22}. For our final solution, 104, 
we decided to adopt the
pole estimate of \citet{scheirich22} 
because it is more precise.
Figure~\ref{fig:solcomp}
compares projections of best-fit parameters and corresponding formal
$3\sigma$ uncertainties of solution 104 with solutions having 
shorter data-arcs, 2003-2019, and
2003-2020. Each successive solution has smaller uncertainties
and they are completely contained within the uncertainties
of previous solutions, indicating that the three 
solutions are consistent and able to predict future solutions.

The formal $3\sigma$ uncertainty in the orbital phase 
of solution 104 during the planned DART
impact date of 2022 September 26 is 
$5.4^\circ$.  The formal uncertainties are underestimated because they 
do not take into account variations 
due to the sizes of the primary and secondary, 
orbit pole, eccentricity, argument of pericenter,
and the obliquity of the orbit pole with 
respect to the spin pole of the primary.
\citet{naidu20a} recommended that
the formal uncertainties derived from the 
covariance matrix be multiplied by 
a factor of 1.3 to capture possible systematic 
errors due to unestimated parameters.
Comparing the uncertainties with those from 
\citet{scheirich22} suggests that
the formal uncertainties in 
Table~\ref{tab:bestfit} might be underestimated by 
up to a factor of 3.

\begin{table*}\footnotesize
  \centering
  \caption{Best-fit orbital parameters of solution 104.
   The solution epoch was chosen to be JD 2455873.0 UTC 
   for easy comparison with \citet{scheirich22}.
   $M_0$, $n_0$ and $\dot{n}$ were estimated. Pole ($\lambda, \beta$) is not
   estimated and is taken from ~\citet{scheirich22}. 
   The osculating period is derived from $n_0$. 
   $GM_{sys}$ is the standard gravitational parameter of the system and is derived from
   the estimated value of $n_0$ and the assumed value of the semimajor axis at epoch.
   $\chi^2_\nu= \chi^2/(n_{obs}-n_{est})$ 
   is the reduced $\chi^2$, where $n_{obs}$ is the number of 
   observations and $n_{est}$ is the number of estimated parameters.}
  \begin{tabular}{lll}
    Parameter                      & Value $\pm~1\sigma$\\
    \hline
    $M_0$ ($^\circ$)            & 89.2 $\pm$ 1.8      \\
    Period (h)                       & 11.9216262 $\pm$ 0.0000027      \\
    $n_0$ (rad s$^{-1}$)      & $(1.46400266 \pm 0.00000035)\times 10^{-4}$  \\
    $\dot{n}$ (rad s$^{-2}$)  & $(5.0 \pm 1.0) \times 10^{-18}$    \\
    Epoch (UTC)                  &  2011-11-07 12:00:00                 \\
    $\chi^2$                          & 21.6                     \\
    $\chi^2_\nu$                   & 0.37                            \\
    $(\lambda, \beta)^\circ$  & $(320.6,-78.6)^\circ$                \\
    $GM_{sys}$ (m$^3$ s$^{-2}$)               & 37.03627274882414 & \\
  \end{tabular}
  \label{tab:bestfit}
\end{table*}

\begin{table}
  \centering
  \caption{Covariance matrix corresponding to 
  solution 104 in Table~\ref{tab:bestfit} at epoch
    2011-11-07 12:00:00 UTC. Units of the
    parameters are in radians and seconds. }

  \begin{tabular}{l|lll}
               & $M_0$                      & $n_0$                 & $\dot{n}$\\\hline
 $M_0$     & $1.03789679 \times 10^{-03}$  & $1.65661906\times 10^{-13}$ & $-3.10150235\times10^{-20}$\\
 $n_0$     & $1.65661906\times10^{-13}$   & $1.13162377\times10^{-21}$ & $-3.31164606\times10^{-30}$ \\
 $\dot{n}$ & $-3.10150235\times10^{-20}$  & $-3.31164606\times10^{-30}$ & $9.93383023\times10^{-37}$ \\
  \end{tabular}\\  
  \label{tab:cov}
\end{table}

\section{Discussion and future work}
We used solution 104 to predict the types and times of mutual events 
around the time of the planned DART impact. We find that there are only eclipses and no
occultations in 2022 September and the formal $3\sigma$ uncertainties 
of the times of these eclipses are about 11 minutes.
However, there are photometric observing opportunities
before the DART impact, starting around June-July 2022.
By performing a covariance analysis and assuming 
two mutual event observations each in July, August, and September, we find that the 
$3\sigma$ uncertainty in orbital phase at the time of the DART impact will improve
modestly, from $5.4^\circ$ to $4.2^\circ$. This corresponds to $3\sigma$ uncertainties in 
mutual event time predictions of about 8 minutes. This uncertainty could be reduced further if 
the secondary-to-primary separation can be measured in spatially
resolved images of the two components taken from the DART spacecraft prior to impact.
Such measurements can be used in the orbit fit.

To estimate the change in the mean motion, $\Delta n$, due to the DART impact, 
we will modify the post-impact orbit model as follows:

\begin{equation}
  \begin{aligned}
  M(t) = M_{imp} + (n_{imp} + \Delta n)(t-t_{imp}) ,\\
  n(t) = n_{imp} + \Delta n,\\
  \end{aligned}
  \label{eq:postimp_model}
\end{equation}
where $t_{imp}$ is the time of impact. $M_{imp}$ and $n_{imp}$ are the mean anomaly 
and mean motion at $t_{imp}$ and are calculated by substituting $t = t_{imp}$ in
equation~\ref{eq:model}. The value of $t_{imp}$ will be available from the DART 
spacecraft navigation team. $\Delta n$ will be treated as a fourth estimable parameter in the fit. 

\begin{figure}
  \centering
  \includegraphics[]{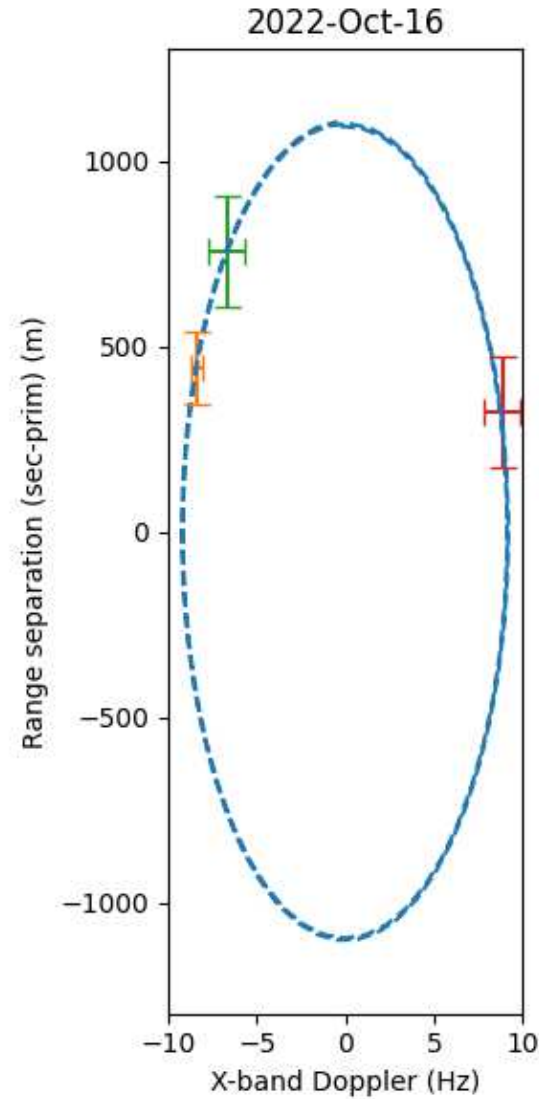}
  \caption{Blue dashed lines show the path of the secondary relative to the primary on 2022 October 16 UTC 
  in range-Doppler space in radar images. Earth is at the bottom and orbital motion is 
  clockwise. The orange point shows the predicted location of Dimorphos
  at 10:00 UTC based on the pre-impact orbit (solution 104). The green and red points show predicted locations of
  Dimorphos assuming a 1- and a 7-minute period change. Error bar on the orange point is its formal $3\sigma$ 
  uncertainties. The error bars on the green and red points are expected 
  measurement uncertainties from Goldstone-Green Bank Telescope range-Doppler images 
  (150 m in range and 1 Hz in Doppler).}
  \label{fig:radpred}
\end{figure}

Studies predict that the impact is expected to reduce the orbital period of Dimorphos by at
least 7 minutes~\citep{cheng16, rivkin21}; the minimum change for mission 
success, a “level 1 requirement,” is at least 73 seconds. We plan on using ground-based 
radar and lightcurve observations to estimate $\Delta n$. 
Radar will provide the first opportunity to detect a change in the orbit with the observing window at Goldstone 
starting on 2022 September 25 \citep{naidu20} and extending through the middle of November. 
Radar range measurements of the secondary relative to the primary
with uncertainties of 150~m are expected between October 2 and 22. Figure~\ref{fig:radpred} shows the drift 
in Dimorphos in range-Doppler space due to a 1-min and a 7-min change in orbit period relative to 
the unperturbed orbit. Such a drift will be detectable in radar images a few weeks after the impact. 

Lightcurve observations are expected for several months after the impact, until around 2023 March.
This will offer a longer time baseline than the radar observations, which should provide tighter 
constraints on $\Delta n$. We will use the mutual event times seen in post-impact lightcurves
along with the radar range and Doppler measurements to estimate $\Delta n$.

\section*{Acknowledgement}
This work was carried out at the Jet Propulsion Laboratory, California Institute of
Technology, under a contract with the National Aeronautics and Space Administration
(80NM0018D0004).
The work by P. Pravec and P. Scheirich was supported by the Grant
Agency of the Czech Republic, Grant 20-04431S.
\\
\\
\copyright 2022. All rights reserved. 

\bibliography{references}{}
\bibliographystyle{aasjournal}



\end{document}